\documentclass[fleqn,usenatbib]{mnras}

\usepackage{newtxtext,newtxmath}

\usepackage[T1]{fontenc}

\DeclareRobustCommand{\VAN}[3]{#2}
\let\VANthebibliography\thebibliography
\def\thebibliography{\DeclareRobustCommand{\VAN}[3]{##3}\VANthebibliography}

\usepackage{graphicx}	
\usepackage{amsmath}	
\usepackage{deluxetable} 

\title[Melnick 39 is a colliding-wind binary]{Melnick 39 is a very massive intermediate-period colliding-wind binary}

\author[A. M. T. Pollock et al.]{
A. M. T. Pollock,$^{1}$\thanks{E-mail: A.M.Pollock@sheffield.ac.uk}
P. A. Crowther,$^{1}$
J. M. Bestenlehner,$^{1}$
Patrick S. Broos$^{2}$
and Leisa K. Townsley$^{2}$\thanks{Published posthumously}
\\
$^{1}$Astrophysics Research Cluster, School of Mathematical and Physical Sciences, University of Sheffield, Hounsfield Road, Sheffield, S3 7RH, UK\\
$^{2}$Department of Physics \& Astronomy, 525 Davey Laboratory, Pennsylvania State University, University Park, PA 16802, USA\\
}

\date{Accepted 2025 March 20. Received 2025 March 20; in original form 2024 July 18}

\pubyear{2025}

\begin{document}
\label{firstpage}
\pagerange{\pageref{firstpage}--\pageref{lastpage}}
\maketitle

\begin{abstract}
Individually identified binary systems of very massive stars define fixed points on possible evolutionary pathways that begin with extreme star formation and end in either coalescence of compact remnants or complete disruption as pair-production supernovae. 
The LMC star Melnick~39 in the Tarantula Nebula is revealed to be an eccentric ($e = 0.618\pm0.014$) binary system of reasonably long period from time-series analysis of {\it Chandra} T-ReX X-ray observations. Its  
X-ray luminosity scales with the inverse of the binary separation, as expected for colliding-wind binaries in the adiabatic regime. The inclusion of optical time-series spectroscopy from the
VLT FLAMES Tarantula Survey and archival {\it HST} spectroscopy confirms Melnick~39 as a double-lined O2.5\,If/WN6+O3\,V--III spectroscopic binary with orbital period near 648 days. We obtain a mass ratio of $q = 0.76 \pm 0.06$, and minimum dynamical masses of $105\pm11$ and $80\pm11 M_{\sun}$ for the O2.5\,If/WN6 and O3\,V-III components, plus photometric evidence for an orbital inclination near 90$^{\circ}$. Disentangled spectroscopy allows the physical and wind properties of the primary to be determined, including $T_{\ast}$ = 44 kK, $\log L/L_{\sun}$ = 6.2, $\log \dot{M}/M_{\sun}$ yr$^{-1}$ = $-5.0$. Its dynamical mass agrees closely with $109 M_{\sun}$ obtained from the mass-luminosity relation of very massive stars.
\end{abstract}




\section{Introduction} \label{intro}

The detection of compact mergers by LIGO/Virgo gravitational wave observatories has led to renewed interest in massive binaries, especially within the low metallicity environments of the Magellanic Clouds, some of which may have been identified as candidate progenitors of black hole mergers \citep{belcynski2022}. It is now established that the majority of massive stars are members of binary systems \citep{sana2012sci, sana2013}. 
Long-period systems can be identified via high spatial resolution imaging or interferometry, while short-period systems are preferentially detected through eclipses or spectroscopic campaigns \citep{MoeDiStefano2017}.

X-ray monitoring offers one route to identifying the more observationally challenging intermediate period systems owing to X-ray variability arising from wind-wind collisions within eccentric orbits. The Tarantula Nebula hosts the richest massive stellar population within the Local Group \citep{crowther2019} and has been observed with the {\it Chandra} X-ray Visional Project `The Tarantula -  Resolved by X rays' designated T-ReX \citep{Townsley2024}. Some of its X-ray light curves have indeed been used to identify candidate binary systems such as Melnick 34 \citep{pollock2018, tehrani2019} and Melnick 33Na \citep{bestenlehner2022} subsequently confirmed as such.

Melnick~39\footnote{SIMBAD: "Cl$\ast$ NGC 2070 MEL 39"} \citep{melnick1985}, hereafter Mk~39 (aka VFTS 482, Brey 78 or BAT99 99 notwithstanding some SIMBAD name confusion) is an Of/WN star \citep{walborn1997, crowther2011}, located at a projected distance of 3 pc NW of R136 at the heart of the Tarantula Nebula. Spectroscopic analysis of Mk~39 by \citet{bestenlehner2014} revealed a very high bolometric luminosity ($\log L/L_{\sun} \sim$ 6.4) and inferred stellar mass ($\sim145 M_{\sun}$). Mk~39 is one of the brightest X-ray sources in the Tarantula Nebula \citep{portegies2002, townsley2006}. \citet{crowther2022} analysed the cumulative T-ReX dataset to determine a mean $\log L_{\rm X}/({\rm erg\,s}^{-1}$) = 34.3, while other tentative evidence for binarity has also been reported \citep{massey2002, massey2005} including a preliminary orbital period of 92.6$\pm$0.3 days from radial velocity variability \citep{schnurr2008}. These collective characteristics are strongly suggestive of a colliding-wind binary system involving very massive components \citep{stevens1992}. 

Large samples of OB stars in the Tarantula Nebula were observed spectroscopically via the VLT/FLAMES Tarantula Survey \citep[VFTS,][]{evans2011}, which included Mk~39, although  follow-up radial velocity studies have been focused on normal OB stars \citep{mahy2020,villasenor2021}. 
In this paper we confirm the binary nature of Mk~39;
establish its orbital period;
and determine its component mass ratio from analysis of VFTS and archival {\it Hubble Space Telescope (HST)} spectroscopy.

\begin{figure}
\includegraphics[width=\columnwidth]{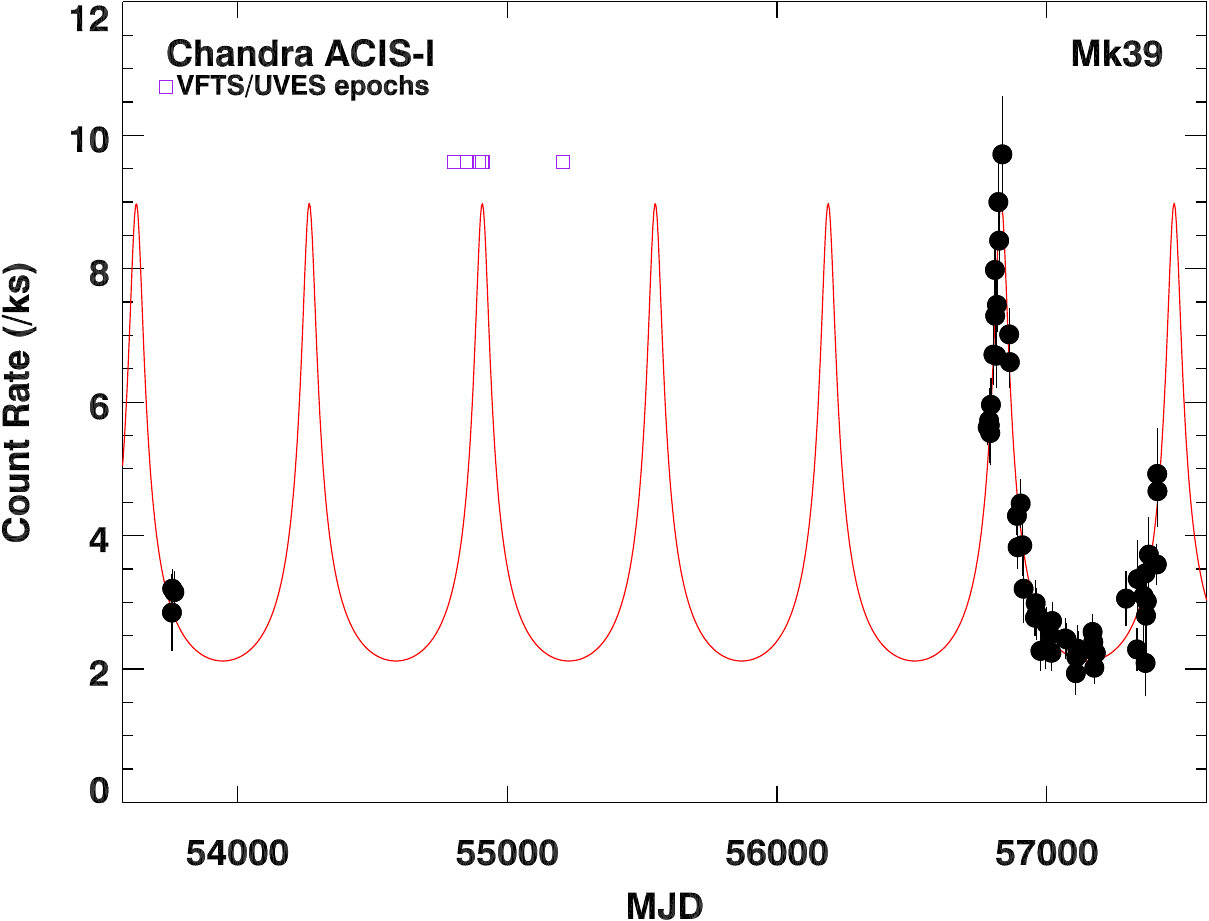}
\caption{Least-squares fit of $1/D$ to {\it Chandra} ACIS-I T-ReX sensitivity-corrected count rates of Mk~39 \ supplemented by earlier ACIS-I data from 2006 \citep{townsley2014} 
implying $P = 641.1_{-3.3}^{+4.3}$ d, $e$ = 0.618$\pm$0.014, $T_0$ = 56830.6$\pm$1.4 MJD. Also shown are the 7 epochs, some overlapping, of the VFTS optical spectroscopy reported in Table~\ref{RV1}}. \label{light_curve} 
\end{figure}

\section{Orbital period of Mk~39 from X-rays}\label{TREX}

The 2Ms {\it Chandra} T-ReX program \citep{Townsley2024} was obtained over 630 days between 2014 May 3 and 2016 January 22 using the ACIS-I instrument centred on R136a, the central cluster of the Tarantula Nebula.
For this study we incorporate 92~ks of ACIS observations from 2006 January 21-30 \citep{townsley2014}. Data reduction, point-source detection and extraction are summarised in \citet{crowther2022} who noted that Mk~39 is an X-ray variable. The X-ray luminosities of colliding-wind binaries in the adiabatic regime are expected to depend on the inverse of the binary separation, $D$, \citep[e.g.][]{stevens1992}
as observed closely to apply over most of the long-period eccentric orbit of WR~140 \citep{Pollock2021}. 
 Fig.~\ref{light_curve} provides a least-squares fit of $1/D$ to the full dataset, revealing an excellent match for an 
 orbital period of $P = 641.1_{-3.3}^{+4.3}$ days, $e = 0.618\pm0.014$ and $T_{0} = 56830.6\pm1.4$ MJD.
 The inferred orbital period narrowly exceeds the length of the 2014--2016 T-ReX campaign, which provides a strict lower limit to the period as the rising portion of the
 light curve at the end of the T-ReX campaign did not reach the level observed at the beginning.
 This also accounts for asymmetric errors in the period estimate.
 Critical in obtaining the solution were individual observation sensitivity corrections and inclusion of the 
 measurements made in 2006. Details of the X-ray solution and an observation log are given in Appendix \ref{X-rayDetails}.

\begin{table}
\caption{Keplerian orbital solution for Mk~39 from fits to the T-ReX X-ray light curve \citep{Townsley2024} and to optical spectroscopy obtained in this study from VFTS \citep{evans2011} and {\it HST} \citep{massey2005}.}
\label{solution}
\begin{center}
\begin{tabular}{lll}
\hline\hline
Parameter & Result & Method \\
\hline
$P$ 			&  648.6 $\pm$ 0.9 d      			& VFTS, {\it HST} \\
$T_{0}$ (MJD) 	& 56830.6 $\pm$ 1.4    			& {\it Chandra} T-ReX \\
$e$               	& 0.618 $\pm$ 0.014           		& {\it Chandra} T-ReX \\ 
$K_1$ 		&   76.9 $\pm$ 4.3 km\,s$^{-1}$ 	& VFTS, {\it HST} \\
$K_2$ 		& 101.4 $\pm$ 6.6 km\,s$^{-1}$ 	& VFTS, {\it HST} \\
$v_{\rm sys}$ 	& 260.5 $\pm$ 3.2 km\,s$^{-1}$ 	& VFTS, {\it HST} \\
$\omega$     	& 130.4 $\pm\ 3.6^{\circ}$  		& VFTS, {\it HST} \\
\hline
\end{tabular}
\end{center}
\end{table}

\begin{figure}
\includegraphics[angle=-90,width=\columnwidth]{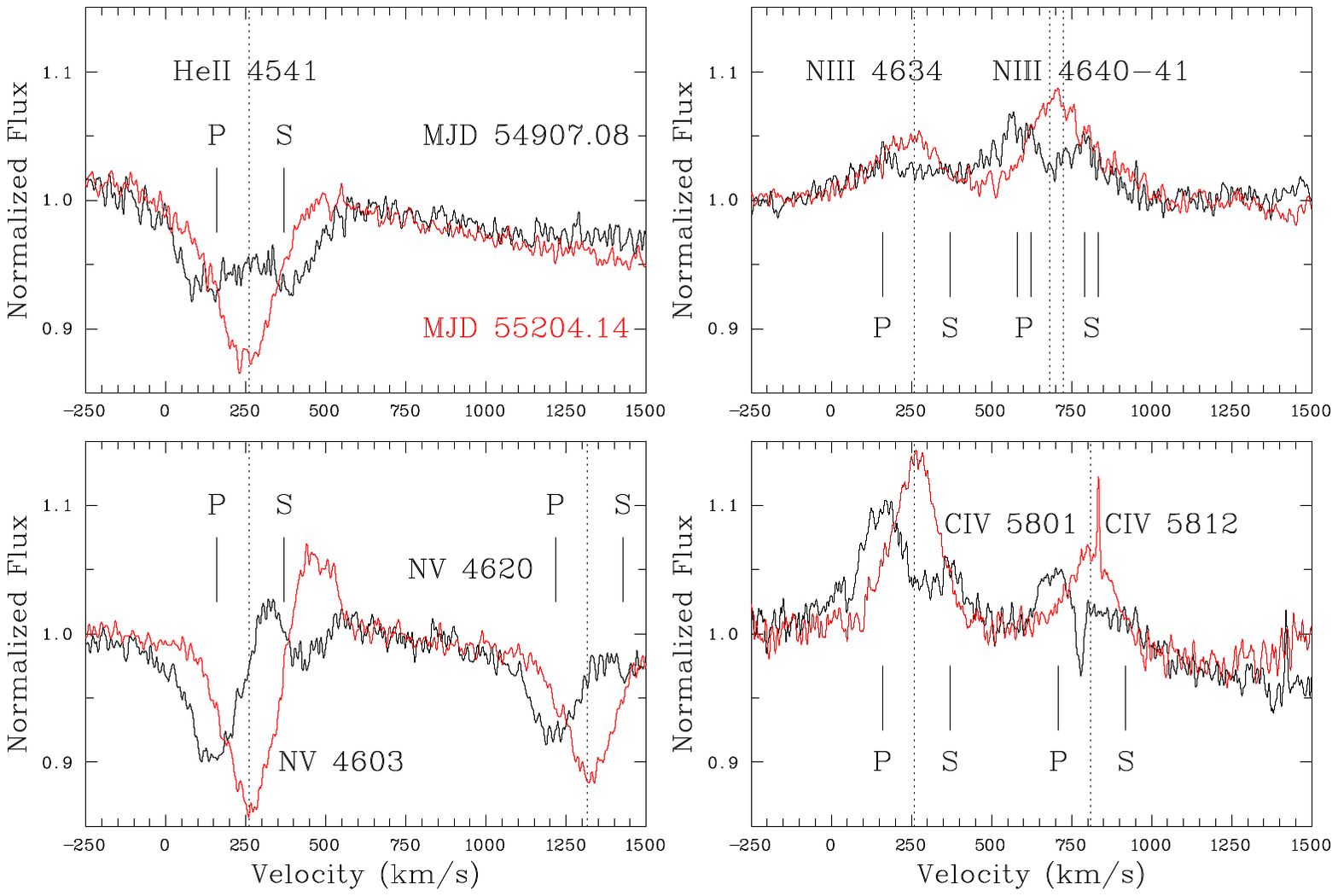}
\caption{Observed VLT/UVES radial velocities of selected lines of Mk~39 (black: MJD 54907.08, red: MJD 55204.14) indicating primary (P) and secondary (S) components and systemic velocities (dotted lines)
\label{uves} }
\end{figure}

\section{Optical orbital solution for Mk~39}\label{VFTS}

In order to assess the reliability of the X-ray orbital solution of Mk~39 (VFTS 482) and establish the nature of the individual components we have
taken advantage of archival spectroscopy
acquired at 7 epochs via the VLT/FLAMES Tarantula Survey \citep{evans2011} between 2008 December 3 and 2010 January 8 (MJD 54803--55204)
with the UV-Visual Echelle Spectrograph \citep[UVES,][]{D'Odorico2000}.
For each epoch two exposures of 1815~sec were obtained with the red arm using the $\lambda$520\,nm central wavelength, providing coverage between 4175--5155\AA\ and 5240--6200\AA\ at a spectral resolution of $R$=53,000. 

UVES reveals spectroscopic variability of Mk~39, including double lines of \ion{He}{II} $\lambda$4542, 5412 from 2009 March 3--17 (MJD 54893--54907), as indicated in Fig.~\ref{uves}, fortuitously corresponding to what proves to be a quadrature phase near X-ray maximum  (Fig.~\ref{light_curve}), supporting the X-ray orbital solution and establishing Mk~39 as a double-lined spectroscopic binary (SB2). \ion{He}{II} $\lambda$4686 is severely blended at all epochs, while the cores of the H$\beta$ and H$\gamma$ absorption lines are contaminated with strong nebular emission. 

We have supplemented VLT/UVES spectroscopy of Mk~39 with the archival {\it HST}/FOS spectroscopy described by \citet{massey2005} from 1997 January 1 (GO 6417, P.I. P.~Massey) and {\it HST}/STIS spectroscopy from 1998 February 4 (GO 7739, P.I. P.~Massey). The FOS dataset used the G400H grating, covering $\lambda\lambda$3235--4781 at $R\sim1300$. The STIS dataset used the G430M/4451 setting, covering $\lambda\lambda$4310--4593 at $R\sim6000$, while the G750M/6581 setting covers $\lambda\lambda$6297--6866 at $R\sim5000$ and includes H$\alpha$. These datasets greatly extend the \ion{He}{II} $\lambda$4542 spectroscopic baseline, and so help to constrain the orbital period of the system.
 
\begin{table*}
\caption{Radial velocities in km\,s$^{-1}$ of spectral features in the primary and secondary components of Mk~39 from VLT/UVES spectroscopy. 
Epochs correspond to the midpoints of observations.}
\label{RV1}
\begin{tabular}{lrrrrrrr}
\hline\hline
MJD & \multicolumn{2}{c}{\ion{He}{II}  $\lambda$4542} 
        & \multicolumn{1}{c}{\ion{N}{V} $\lambda$4603} 
        & \multicolumn{1}{c}{\ion{N}{III} $\lambda$4641} 
        & \multicolumn{2}{c}{\ion{He}{II}  $\lambda$5412} 
        & \multicolumn{1}{c}{\ion{C}{IV} $\lambda$5801} \\
        & \multicolumn{1}{c}{Primary} & \multicolumn{1}{c}{Secondary} 
        & \multicolumn{1}{c}{Primary} 
        & \multicolumn{1}{c}{Primary} 
        & \multicolumn{1}{c}{Primary} & \multicolumn{1}{c}{Secondary} 
        & \multicolumn{1}{c}{Primary} \\
\hline
54803.17             &   290.9$_{-3.3}^{+3.0}$ & 210.1$_{-3.4}^{+3.2}$ & 299.2$_{-3.8}^{+3.9}$ & 248.7$_{-5.1}^{+5.1}$ & 305.0$_{-4.0}^{+3.9}$ &  188.3$_{-4.1}^{+3.8}$   & 291.7$_{-4.7}^{+4.6}$     \\ 
54803.22             &   292.3$_{-2.8}^{+2.8}$ & 205.1$_{-3.8}^{+2.5}$ & 300.7$_{-3.7}^{+3.7}$ & 240.4$_{-4.2}^{+4.1}$ & 301.1$_{-3.6}^{+3.6}$ & 185.4$_{-3.6}^{+3.6}$    &  290.5$_{-4.5}^{+4.5}$  \\
54851.10             & 295.1$_{-4.3}^{+2.2}$ & 215.9$_{-4.2}^{+4.4}$  & 283.3$_{-3.5}^{+3.5}$ & 238.3$_{-5.3}^{+5.3}$ & 314.5$_{-3.6}^{+3.6}$ & 182.8$_{-3.6}^{+4.0}$      & 282.5$_{-4.7}^{+4.7}$ \\
54893.04             & 105.9$_{-4.1}^{+4.8}$ & 401.7$_{-3.8}^{+3.8}$  & 157.8$_{-3.9}^{+3.9}$ & 95.5$_{-7.0}^{+7.0}$ & 114.1$_{-4.2}^{+4.5}$ & 398.4$_{-5.1}^{+5.4}$          & 171.0$_{-4.2}^{+4.2}$ \\
54906.01            &  152.6$_{-3.9}^{+3.4}$ & 379.6$_{-7.2}^{+7.4}$  & 186.1$_{-4.8}^{+4.8}$ & 133.8$_{-9.1}^{+9.0}$ & 113.4$_{-5.6}^{+5.7}$ & 378.3$_{-5.8}^{+5.8}$       & 187.2$_{-5.8}^{+5.6}$ \\
54907.08            &  132.6$_{-4.7}^{+4.8}$ & 396.4$_{-4.2}^{+5.9}$  & 177.9$_{-3.3}^{+3.3}$ & 135.1$_{-4.7}^{+4.7}$ & 138.4$_{-3.7}^{+3.8}$ & 364.8$_{-4.2}^{+4.2}$        &  198.3$_{-4.5}^{+4.5}$\\
55204.14            &  294.2$_{-3.0}^{+3.2}$ & 215.7$_{-3.9}^{+3.8}$  & 297.1$_{-3.7}^{+3.7}$ & 253.9$_{-4.1}^{+4.1}$ & 298.5$_{-3.0}^{+3.2}$ & 195.3$_{-3.9}^{+3.8}$        & 293.2$_{-4.4}^{+4.4}$ \\
\hline
\end{tabular}
\end{table*}

Mk~39 has received previous spectral classifications of O4\,If \citep{melnick1985}, O3\,If*/WN6 \citep{walborn1997}, O2\,If* \citep{massey2005} and O2.5\,If/WN6 \citep{crowther2011}, the latter based on the UVES data used here.  The primary is confirmed as an O2.5\,If/WN6 star, since the morphology of H$\beta$ is a clear P Cygni profile. The UVES observations close to periastron (Fig.~\ref{uves}) permit the spectral type of the secondary to be determined. From comparison with early-O templates \citep{walborn2002},  \ion{He}{II} $\lambda$4542, 5412 are strong, with \ion{He}{I} $\lambda$4471 weak or absent, plus weak \ion{N}{V} $\lambda$4603--20 absorption and \ion{N}{III} $\lambda\lambda$4634-41 emission, implying O3\,V-III for the secondary. 

We have undertaken single or double gaussian fits to various optical absorption lines in VLT/UVES spectra to establish the individual component radial velocities presented in Table~\ref{RV1}. In addition, we have used a grid of {\sc cmfgen} model atmospheres \citep{hillier1998} suitable for early-type O stars of LMC metallicity \citep{bestenlehner2014} to cross correlate with VLT/UVES and {\it HST}/FOS+STIS spectroscopy. Radial velocities of the primary and secondary components are presented in Table~\ref{RV2}, having been derived by investigating the extrema and zero-points of the second and third derivatives of the cross-correlation function. 

\begin{figure*}
\includegraphics[width=\textwidth]{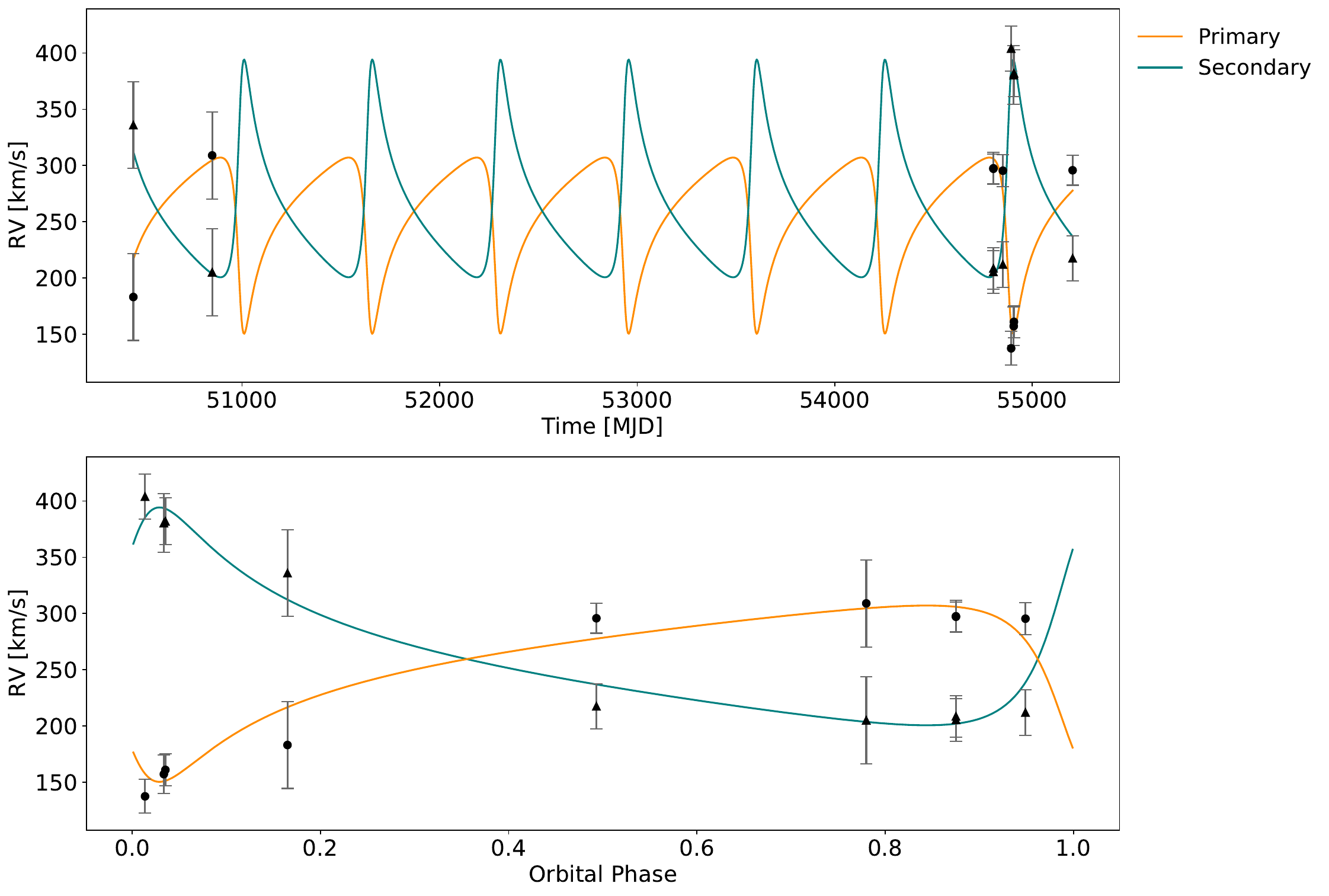}
\caption{Orbital solution to \ion{He}{II} $\lambda$4542 VLT (UVES) and {\it HST} (FOS+STIS) radial velocities for the primary (filled circles) and secondary (filled triangles) components of Mk~39, revealing $K_{1}$ =  76.9 km\,s$^{-1}$ and $K_{2}$ = 101.4 km\,s$^{-1}$, $v_{\rm sys}$ = 260$\pm$3 km\,s$^{-1}$, $P$ = 648.6$\pm$0.9 days for $e$ = 0.618 obtained from X-ray light curve fit, implying a mass ratio of $q = 0.76 \pm 0.06$. \label{orbit} }
\end{figure*}

To derive the orbital parameters we have used these uncertainty weighted averages and employed the MCMC approach of \citet{tehrani2019}. As the VLT/UVES data provide poor orbital phase coverage, we fixed the eccentricity to the X-ray photometric solution from Section~\ref{TREX} and utilised $T_{0}$ = 56830.6$\pm$1.4 MJD as a prior.
We obtained the solution reported in Table~\ref{solution} and illustrated in Figure~\ref{orbit}
with period $P = 648.6\pm0.9$~d, leading to
minimum masses $M_{1} \sin^3 i = 105\pm11 M_{\sun}$ and  $M_{2} \sin^3 i = 80\pm11 M_{\sun}$
and mass ratio $q = K_1/K_2 = M_{2}/M_{1} = 0.76 \pm 0.06$.
The systemic velocity $v_{\rm sys}$ = 260$\pm$3 km\,s$^{-1}$ of Mk~39 is close to the mean radial velocity of 268$\pm$6 km\,s$^{-1}$ of OB stars within 5 pc of R136 
\citep{henault2012}.

The X-ray and optical period estimates differ by between $1$ and $2\sigma$ due to larger positive errors in the X-ray value with any inconsistency probably reflecting systematic errors of which no account has been taken. Appendix~\ref{X-rayDetails} below discusses some of these in the X-ray 1/D regime. In the optical, the solution is quite well constrained by the appearance of double lines in 3 epochs over an interval of 14 days with the closest of the 4 epochs in which the double lines were not resolved in a key observation 42 days earlier. An alternative solution to that reported in Table~\ref{solution}
is available by exchanging primary and secondary radial velocities in that key observation at MJD 54851.10. Although this is formally superior in statistical terms, the implied period of $P$ = 653.2$\pm$0.4~d is less consistent with the X-ray value. These uncertainties should be eliminated by future suitably well-timed observations in either X-ray photometry or optical spectroscopy
around periastron the next of which is expected to occur in late 2026.

\begin{table}
\caption{Radial velocities in km\,s$^{-1}$ of primary and secondary components of Mk~39 from cross-correlation of VLT/UVES and {\it HST}/FOS+STIS spectroscopy.
Epochs correspond to the midpoints of observations.}
\label{RV2}
\begin{center}
\begin{tabular}{lccc}
\hline\hline
\multicolumn{1}{c}{MJD} & \multicolumn{1}{c}{Dataset} &  \multicolumn{1}{c}{Primary} & \multicolumn{1}{c}{Secondary} \\
\hline
50449.89             & HST/FOS & 183$\pm$10 & 336$\pm$10 \\
50848.83             & HST/STIS & 309$\pm$10 & 205$\pm$10 \\
54803.17             & VLT/UVES & 315$\pm$7 & 195$\pm$7 \\
54803.22             & VLT/UVES & 315$\pm$7 & 196$\pm$7 \\ 
54851.10             & VLT/UVES & 303$\pm$7 & 225$\pm$7 \\
54893.04             & VLT/UVES & 125$\pm$6 & 393$\pm$5 \\ 
54906.01            &  VLT/UVES & 140$\pm$6 & 366$\pm$5 \\ 
54907.08            &  VLT/UVES & 147$\pm$5 & 361$\pm$5 \\
55204.14            &  VLT/UVES & 288$\pm$9 & 226$\pm$8 \\ 
\hline
\end{tabular}
\end{center}
\end{table}

\section{Physical and wind properties of Mk~39}\label{properties}

Armed with the orbital solution of Mk~39 from the previous section, we have disentangled the components of the UVES spectroscopy following the approach of \citet{bestenlehner2022}
for the SB2 system Mk~33Na, which includes rescaling according to the flux ratio of the individual components. For the physical and wind properties of the individual components of Mk~39. \citet{bestenlehner2014} obtained a system luminosity of $\log L/L_{\sun}$ = 6.4 that we adopt here ($E(B-V) = 0.45, R_{\rm V} = 3.26$), together with the mass ratio $q = 0.76$. For LMC very massive main sequence stars   $L \propto M^{1.4 \pm 0.2}$  \citep{kohler2015}, so $L_{2}/L_{1} \sim 0.7$ and $\log (L_1/L_{\sun}) = 6.2$ and $\log (L_2/L_{\sun}) =  6.0$.
 
\begin{figure*}
\begin{tabular}{cc}
\includegraphics[width=\columnwidth]{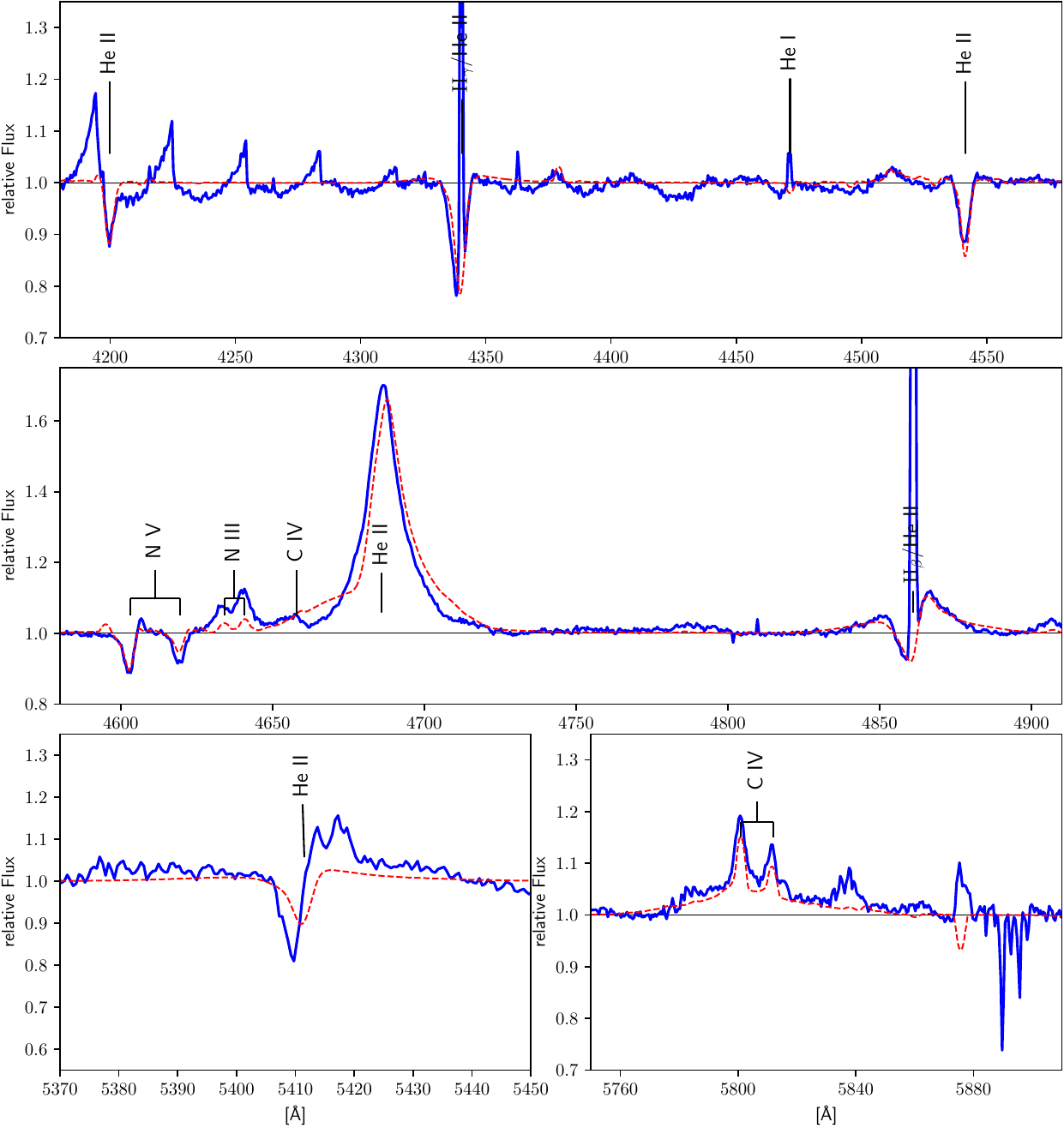} & \includegraphics[width=\columnwidth]{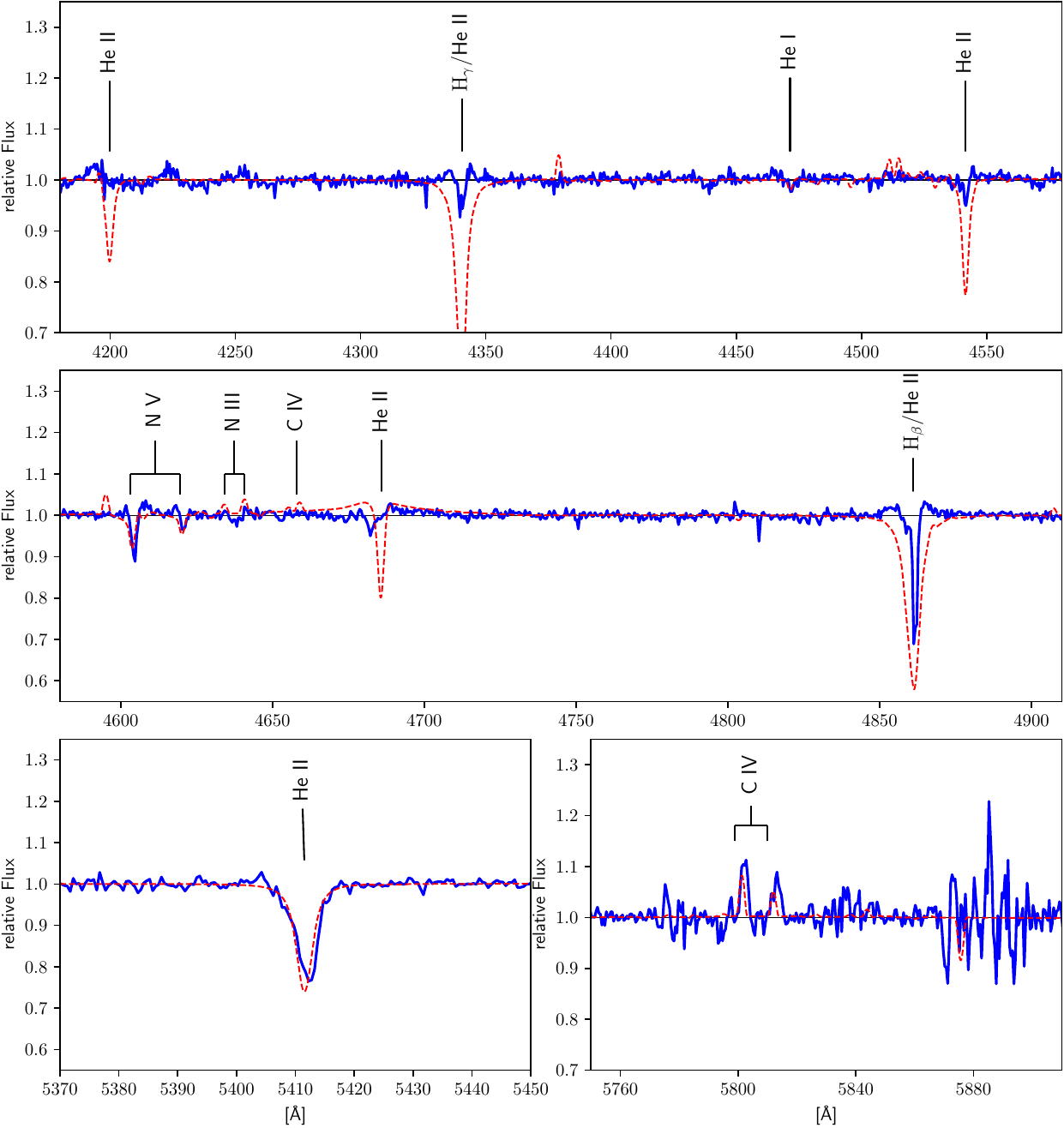} \\
\end{tabular}
\caption{Comparison between disentangled Mk~39 primary (left) and secondary (right) UVES spectroscopy (blue solid lines) and synthetic {\sc cmfgen} spectra (red dotted lines). Strong nebular lines (Balmer, \ion{He}{I} and [\ion{O}{III}])  are present in the left panel, together with instrumental emission features shortward of $\lambda$4300.\label{fits} }
\end{figure*}

\begin{figure}
\begin{center}
\includegraphics[width=\columnwidth]{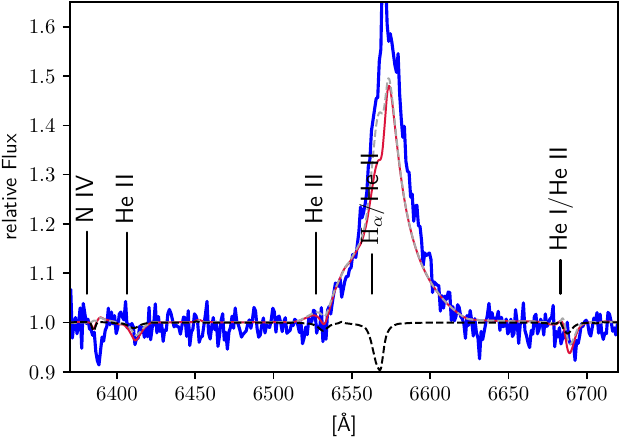}
\end{center}
\caption{HST/STIS spectroscopy of Mk~39 near H$\alpha$ (solid blue) arising from a combination of dominant primary emission and weak secondary absorption. \label{Halpha} }
\end{figure}

 Comparisons between disentangled UVES spectra (blue) and synthetic {\sc cmfgen} spectra (red) for each component are presented in Fig.~\ref{fits} with the STIS H$\alpha$ region in Fig.~\ref{Halpha}. Disentangled secondary spectra at H$\gamma$ and \ion{He}{II} $\lambda$4686 are unreliable owing to the dominant primary spectral lines as well as nebular Balmer-line contamination. We obtain stellar temperatures of $T_{\ast}$ = 44.0$\pm$2.5 kK and 48.0$\pm$2.5 kK respectively for the primary and secondary from \ion{N}{V} $\lambda\lambda$4603-20 and \ion{He}{II} $\lambda$4542/5411 since \ion{He}{I} $\lambda$4471, $\lambda$5876 are dominated by nebular emission. \ion{N}{IV} $\lambda$4058 is not included in the UVES data, so use of solely nitrogen diagnostics would rely on the \ion{N}{V} doublet and \ion{N}{III} $\lambda$4634--41, requiring a lower temperature for the primary and higher temperature for the secondary, subject to uncertainties in their nitrogen abundances. Weak emission in the
 \ion{C}{IV} $\lambda\lambda$5801-12 doublet is seen in both components, and is reasonably well reproduced for our preferred stellar temperatures.
 
 We favour a mass-loss
 rate of about $10^{-5} M_{\sun} {\rm yr}^{-1}$  for the primary from \ion{He}{II} $\lambda$4686, supported by the morphology of H$\beta$ and the STIS H$\alpha$ spectroscopy, whereas $10^{-6.2} M_{\sun} {\rm yr}^{-1}$ is estimated for the secondary owing to an absence of any suitable wind diagnostics. A wind velocity of 2600 km\,s$^{-1}$ is adopted for both components following \citet{bestenlehner2014} who analysed archival far-UV STIS/G140L spectroscopy of Mk~39 from \citet{massey2005}. We estimate
 equatorial rotational velocities
 $v_{\rm eq} \sin i \sim$ 100 km\,s$^{-1}$ and 80 km\,s$^{-1}$ from \ion{He}{II} $\lambda$4542, 5412 lines. 
 
 High-resolution far ultraviolet  spectroscopy of Mk~39 has been obtained via the {\it HST} ULLYSES survey of massive stars in the Magellanic Clouds \citep{RomanDuval2020}, using COS G130M/1291 and G160M/1611, providing
 spectral coverage between $\lambda\lambda$1131--1790 at a spectral resolution of $R\sim$15,000. These datasets were not used for our analysis, but the synthetic spectra provide a
 good match to \ion{O}{V} $\lambda$1371, \ion{Si}{IV} $\lambda\lambda$1393-1402, \ion{He}{II} $\lambda$1640 and \ion{N}{IV} $\lambda$1718, while P Cygni emission for 
 \ion{C}{IV} $\lambda\lambda$1548-51 and \ion{N}{V} $\lambda\lambda$1238-42 is underestimated, the latter mitigated by strong Ly $\alpha$ interstellar absorption and its sensitivity to X-ray production (excluded from our spectroscopic analysis). The reddened spectral energy distribution of Mk~39 provides a good match to COS far-UV, plus more recent STIS near-UV (G230LB) and blue visual (G430L) spectroscopy (GO 16230, P.I. D~Massa). 
 
\begin{table}
\caption{Physical properties of the primary and secondary components of Mk~39 from the present study.}
\label{cmfgen}
\begin{tabular}{llll}
\hline\hline
\multicolumn{1}{c}{Property} &  \multicolumn{1}{c}{Primary} & \multicolumn{1}{c}{Secondary} & \multicolumn{1}{c}{Reference}  \\
\hline
Spectral Type & O2.5\,If/WN6 & O3\,V-III  & This study\\
$T_{\rm eff}$ (kK)           &  44$\pm$2.5 & 48$\pm$2.5 & This study\\
$\log (L/L_{\sun}$)      & 6.20$\pm$0.15 & 6.00$\pm$0.15 & This study \\
$\log \dot{M}/M_{\sun} {\rm yr}^{-1}$ & --5.0$\pm$0.2 & --6.2$_{-0.5}^{+0.2}$ & This study \\
$v_{\infty}$ (km\,s$^{-1}$)  & 2600 & 2600: & \citet{bestenlehner2014} \\
$Y$                             & 0.30$\pm$0.05 & 0.27$\pm$0.05 & This study \\
$v_{\rm eq} \sin i$ (km\,s$^{-1}$) & 100 & 80 & This study \\
$M_{\rm dyn} \sin i$ ($M_{\sun}$)            & 105.2 & 79.8 & This study \\
$M (M_{\sun})$                                & 109$\pm$7 & 83$\pm$5 & \citet{Grafener2011} \\
$M_{\rm evol}$ ($M_{\sun}$)           &    83$_{-18}^{+20}$      &    $69_{-11}^{+14}$     & BONNSAI \\
$\tau$ (Myr)                                      & 1.5$\pm$0.3 &  1.1$_{-0.8}^{+0.3}$ & BONNSAI  \\
\hline
\end{tabular}
\end{table}

BONNSAI\footnote{The BONNSAI web-service is available at \url{www.astro.uni-bonn.de/stars/bonnsai}} \citep{schneider2014} coupled to evolutionary models from   \citet{kohler2015}  and empirical initial rotation velocities of \citet{ramirez-agudelo2013} infers stellar masses and ages of  $M_{\rm evol} = 83_{-18}^{+20} M_{\sun}$ and $69_{-11}^{+14} M_{\sun}$ and  1.5$\pm$0.3 Myr and $1.1_{-0.8}^{+0.3}$ Myr for the primary and secondary, assuming the stars have evolved independently to date. These evolutionary mass estimates are somewhat lower than minimum dynamical mass determinations (BONNSAI underpredicts stellar luminosities by 0.1 dex). The favoured age is consequently $\sim$1.4 Myr, similar to the nearby massive star cluster R136 \citep{crowther2016, brands2022}. 
Alternatively, $M \sim 109 \pm 7 M_{\sun}$ and $83 \pm 5 M_{\sun}$ are obtained for the primary and secondary from the mass-luminosity relationship for very massive stars  \citep{Grafener2011}, suggesting an inclination close to $i = 90^{\circ}$.

\citet[their Fig. 5]{massey2002} reported a short-lived $\sim0.1$ magnitude photometric dip in the optical light curve of Mk~39 \citep[alias HSH 7,][]{hunter1995}
at MJD 51635.36-51.
Our preferred orbital solution in Table~\ref{solution} would suggest a conjunction with the primary in front
at MJD 51627.5$\pm$7.5 with most of the uncertainty due to the orbital timing of period and periastron passage rather than the
orbital geometry of eccentricity and longitude of periastron.
Thus a period shorter within the error budget by 1.0~d would bring consistency between the orbital solution and the occurrence
of a photometric dip due to a physical or wind eclipse that 
is clearly subject to confirmation to reinforce the suggestion of a high orbital inclination.

\section{Discussion and conclusions}\label{summary}

\begin{figure}
\includegraphics[angle=-90,width=\columnwidth]{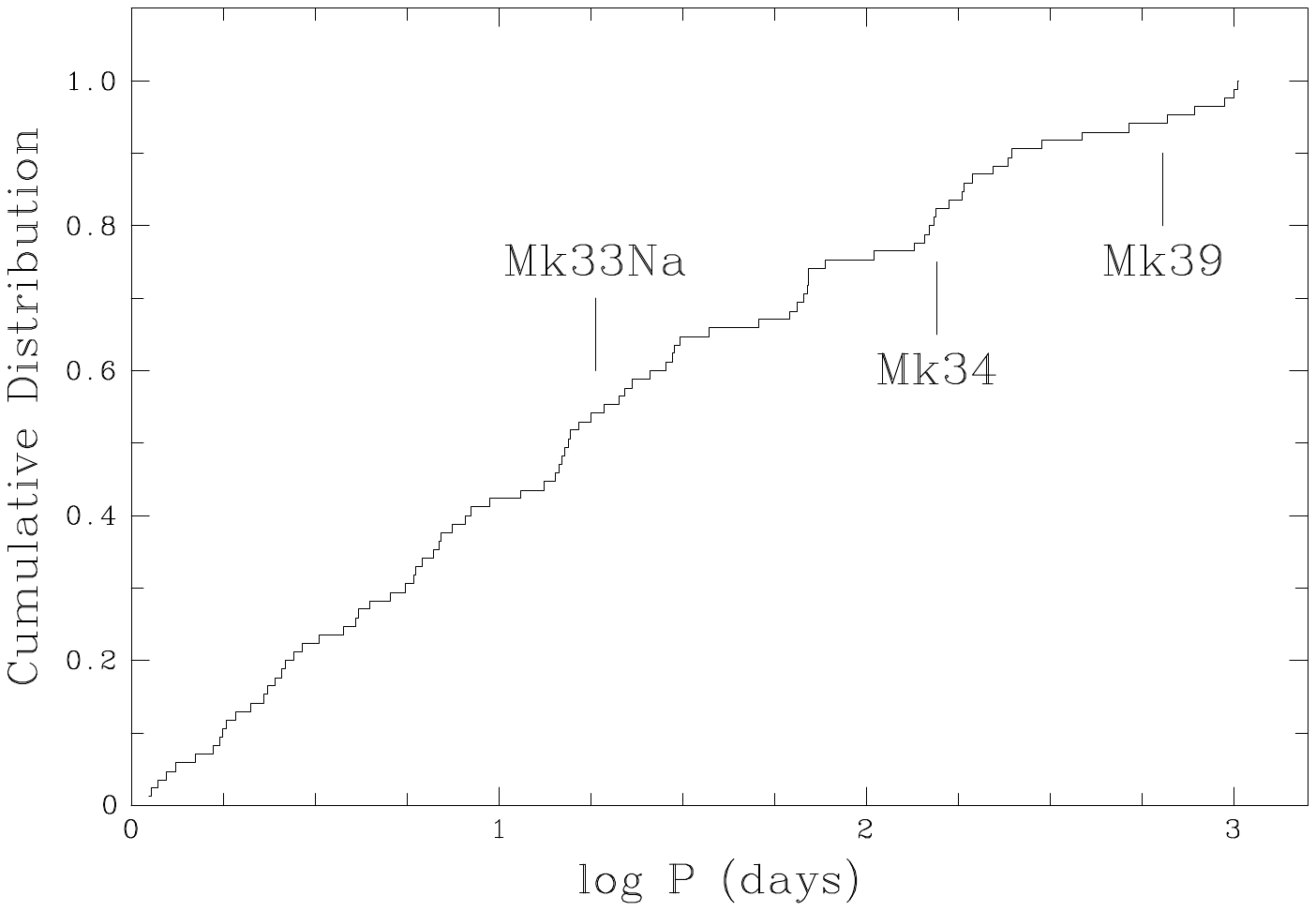}
\caption{Cumulative distribution of orbital periods of OB stars in the Tarantula Nebula from \citet{almeida2017} supplemented with T-ReX SB2 systems, Mk~34 \citep{tehrani2019}, Mk~33Na \citep{bestenlehner2022} and Mk~39 from the present study with some long-period binaries are subject to revision. \label{VFTS-Orbits} }
\end{figure}

We have established that the X-ray luminous Of/WN star Mk~39 in the Tarantula nebula is an SB2 colliding-wind binary with period of 648.6$\pm$0.9 days that follows the inverse separation law for adiabatic emission. In this respect, Mk~39 represents an ideal example with
little or
none of the deviations from adiabatic behaviour seen in other systems near periastron due, for example, to competitive cooling seen in WR~140 \citep{Pollock2021}; 
circumstellar absorption in WR~25, another very massive binary system \citep{Pradhan2021};
or perhaps a combination of both in Mk~34 and WR~21a \citep{pollock2018}.

Mk~39, which reaches a maximum count rate near 10 cts/ks at the minimum periastron orbital separation of 3.5 AU according to the new orbital solution, may be compared with Mk~34, the brightest T-ReX colliding-wind binary at its minimum near 35 cts/ks \citep{pollock2018} at the similar but slightly larger separation maximum of 4.1 AU reached at apastron \citep{tehrani2019}. This difference is roughly consistent with expectations of scaling laws \citep{Luo1990, stevens1992} which may be recast from mass-loss rate and terminal velocity to use luminosity and velocity FWHM, ${\Delta}v$, of the \ion{He}{II} $\lambda4686$ wind
line \citep{Crowther2023} to suggest
$L_X \propto (L{\Delta}v^{-3.2})_{\mathrm{HeII}}$:
X-rays from the weaker, faster wind of Mk~39 fall a factor of a few short of the stronger, slower wind of Mk~34 .

Optical spectroscopy has been used to determine a mass ratio of $q = 0.76 \pm 0.06$, and minimum component masses of $105\pm11 M_{\sun}$ (O2.5\,If/WN6) and $80\pm11 M_{\sun}$ (O3\,V--III). These agree closely with masses from the
 \citet{Grafener2011} mass-luminosity relation for very massive stars based on physical properties determined from disentangled UVES spectroscopy, for an inclination close to 90$^{\circ}$.
The inability of \citet{massey2005} to obtain a satisfactory spectroscopic fit for Mk~39 was attributed to its composite nature though \citet{bestenlehner2014} obtained physical and wind properties of Mk~39 from VLT/UVES spectroscopy \citep{evans2011}, supplemented by {\it HST}/STIS G140L and H$\alpha$ spectroscopy \citep{massey2005} plus K-band VLT/SINFONI spectroscopy. 

The only previously published orbital solution was from \citet{schnurr2008} who obtained $P$ = 92.6$\pm$0.3 days,  $K_1$ = 91$\pm$19 km\,s$^{-1}$ and $v_{\rm sys}$ = 337$\pm$16 km\,s$^{-1}$ from spectroscopic observations of \ion{He}{II} $\lambda$4686, {\it assuming} - with explicit caution - a circular orbit.
Although narrower photospheric absorption lines, such as \ion{He}{II} $\lambda$4542 used in this study, are more straightforward diagnostics of Keplerian orbital motion than broader, slightly irregular, wind emission lines such as 
\ion{He}{II} $\lambda$4686 used by \citet{schnurr2008},
the radial velocities of this emission line in the current VLT/UVES data follow reasonably closely the primary's absorption lines at roughly the same positive displacement of about 80 km\,s$^{-1}$.
The inconsistency with the new orbital solution must lie elsewhere.
According to the new ephemeris, it is plausible that the minimum and maximum radial velocity excursions that drive the earlier orbital solution coincide with quadratures near a projected
periastron passage of the new eccentric solution on 2002 January 16 during the first season of observations, with the second season encompassing the subsequent apastron. However, this possibility fails because
the maximum and minimum obtained during successive runs 20 days apart in 2001 December are too close together in time and too far apart in velocity to be consistent with the parameters in Table~\ref{solution}.

\citet[][their table~5]{bestenlehner2022} provide a summary of massive binaries in the LMC. Of these, the most extreme systems are exclusive to the Tarantula Nebula: Mk~34 \citep[WN5h+WN5h,][]{tehrani2019}, R144 \citep[WN5--6h+WN6--7h,][]{shenar2021}, R139 \citep[O6.5\,Iafc+O6\,Iaf,][]{mahy2020} and Mk~33Na \citep[OC2.5\,If+O4V,][]{bestenlehner2022} with eccentric orbits in the range 18--155 days, and primary masses comfortably exceeding the previous LMC record holder \citep{massey2002}. 
 
 Mk~39 has similar stellar components, albeit with a significantly longer period orbit, as indicated in Fig.~\ref{VFTS-Orbits} which shows a cumulative distribution of OB orbital periods for the Tarantula Nebula from \citet{almeida2017}, supplemented by T-ReX results. X-ray photometric surveys can help facilitate an improved characterisation of orbital periods, eccentricities and mass ratios of massive stars in the LMC, together with continuing spectroscopic surveys \citep{mahy2020,villasenor2021} at longer wavelengths.

\section*{Acknowledgements}

This is part of a collection of papers publishing posthumously the unfinished work of LWK, the principal investigator of T-ReX. This work was supported by the Chandra X-ray Observatory General Observer grants GO5-6080X (PI: L. Townsley) and by GO4-15131X (PI: L. Townsley) and by the Penn State ACIS Instrument Team Contract SV4-74108. All of these were issued by the Chandra X-ray Center, which is operated by the Smithsonian Astrophysical Observatory for and on behalf of NASA under contract NAS8-03060.  Based in part on observations obtained with the NASA/ESA Hubble Space Telescope, retrieved from the Mikulski Archive for Space Telescopes (MAST) at the STScI. STScI is operated by the Association of Universities for Research in Astronomy, Inc. under NASA contract NAS 5-26555. PAC and JMB are supported by the Science and Technology Facilities Council research grant ST/V000853/1 (PI. V. Dhillon).

\section*{Data Availability}

All the observational data used in this article are freely available as follows:
 the X-ray data through the Chandra Data Archive at \url{https://cxc.cfa.harvard.edu/cda/};
 the optical VLT/FLAMES spectra through the ESO Archive Science Portal at \url{https://archive.eso.org/scienceportal/home};
 and
 the optical HST data through the Mikulski Archive for Space Telescopes at \url{https://mast.stsci.edu/portal/Mashup/Clients/Mast/Portal.html}.
 The synthetic {\sc cmfgen} model atmospheres of the primary and secondary stars are available on request.



\bibliographystyle{mnras}
\bibliography{reference} 




\appendix

\section{The X-ray orbital solution}\label{X-rayDetails}

\begin{table*}
\centering
\caption{Binary-phase-ordered log of Chandra~observations in the T-ReX survey of 30~Doradus and three earlier observations with
observation ID and epoch;
exposure time, T;
sensitivity factor relative to the ObsID 7264 maximum;
sensitivity-corrected count rate per 1000s of Mk~39;
and
phase interval covered, $\phi$, of the 641.1-day best-fit X-ray orbit centred on MJD 56830.6.}
\label{ChandraLog}
\begin{tabular}{rccrrr@{ $\pm$ }lr@{ $\pm$ }l}
\hline
ObsID & date & MJD & T(s) & factor & \multicolumn{2}{c}{Mk~39 (cts/ks)} & \multicolumn{2}{c}{$\phi_{641.1}$(d)} \\
\hline
16445 & 2015-05-27T00:18:12 & 57169.013 & 49310 & 0.940 &   2.6 & 0.3 & $-$302.4 & 0.3 \\
17660 & 2015-05-29T14:55:28 & 57171.622 & 38956 & 0.941 &   2.4 & 0.3 & $-$299.9 & 0.2 \\
16446 & 2015-06-02T11:50:14 & 57175.493 & 47547 & 0.940 &   2.0 & 0.2 & $-$295.9 & 0.3 \\
17642 & 2015-06-08T05:11:14 & 57181.216 & 34438 & 0.915 &   2.2 & 0.3 & $-$290.3 & 0.2 \\
16449 & 2015-09-28T05:35:14 & 57293.233 & 24628 & 0.929 &   3.1 & 0.4 & $-$178.3 & 0.2 \\
18672 & 2015-11-08T01:04:22 & 57334.045 & 30574 & 0.928 &   2.3 & 0.3 & $-$137.5 & 0.2 \\
18706 & 2015-11-10T17:09:59 & 57336.715 & 14776 & 0.925 &   3.3 & 0.6 & $-$134.9 & 0.1 \\
18720 & 2015-12-02T10:49:02 & 57358.451 &  9832 & 0.928 &   3.1 & 0.7 & $-$113.2 & 0.1 \\
18721 & 2015-12-08T17:13:14 & 57364.718 & 25598 & 0.926 &   3.4 & 0.4 & $-$106.8 & 0.2 \\
17603 & 2015-12-09T15:27:36 & 57365.644 & 13778 & 0.929 &   2.1 & 0.5 & $-$106.0 & 0.1 \\
18722 & 2015-12-11T09:09:39 & 57367.382 &  9826 & 0.929 &   2.8 & 0.7 & $-$104.3 & 0.1 \\
18671 & 2015-12-13T23:41:12 & 57369.987 & 25617 & 0.927 &   3.0 & 0.4 & $-$101.6 & 0.2 \\
18729 & 2015-12-21T22:10:30 & 57377.924 & 16742 & 0.929 &   3.7 & 0.6 & $-$93.7 & 0.1 \\
18750 & 2016-01-20T00:41:30 & 57407.029 & 48318 & 0.926 &   3.6 & 0.3 &  $-$64.4 & 0.3 \\
18670 & 2016-01-21T20:59:37 & 57408.875 & 14565 & 0.928 &   4.9 & 0.7 &  $-$62.8 & 0.1 \\
18749 & 2016-01-22T16:14:19 & 57409.677 & 22153 & 0.926 &   4.7 & 0.5 &  $-$61.9 & 0.1 \\
16192 & 2014-05-03T04:10:27 & 56780.174 & 93761 & 0.932 &   5.6 & 0.3 &  $-$49.9 & 0.6 \\
16193 & 2014-05-08T10:15:25 & 56785.427 & 75994 & 0.929 &   5.7 & 0.3 &  $-$44.7 & 0.5 \\
16612 & 2014-05-11T02:15:31 & 56788.094 & 22672 & 0.955 &   5.7 & 0.6 &  $-$42.3 & 0.2 \\
16194 & 2014-05-12T20:00:24 & 56789.834 & 31333 & 0.898 &   5.5 & 0.5 &  $-$40.6 & 0.2 \\
16615 & 2014-05-15T08:24:45 & 56792.351 & 45170 & 0.954 &   6.0 & 0.4 &  $-$38.0 & 0.3 \\
16195 & 2014-05-24T14:09:28 & 56801.590 & 44405 & 0.827 &   6.7 & 0.5 &  $-$28.7 & 0.3 \\
16196 & 2014-05-30T00:05:56 & 56807.004 & 67109 & 0.924 &   8.0 & 0.4 &  $-$23.2 & 0.4 \\
16617 & 2014-05-31T01:27:04 & 56808.060 & 58860 & 0.749 &   7.3 & 0.4 &  $-$22.2 & 0.4 \\
16616 & 2014-06-03T22:26:17 & 56811.935 & 34530 & 0.953 &   6.7 & 0.5 &  $-$18.4 & 0.2 \\
16197 & 2014-06-06T12:32:26 & 56814.523 & 67790 & 0.735 &   7.5 & 0.4 &  $-$15.7 & 0.4 \\
16198 & 2014-06-11T20:20:49 & 56819.848 & 39465 & 0.688 &   9.0 & 0.6 &  $-$10.5 & 0.2 \\
16621 & 2014-06-14T14:46:41 & 56822.616 & 44400 & 0.758 &   8.4 & 0.5 &   $-$7.7 & 0.3 \\
16200 & 2014-06-26T20:01:47 & 56834.835 & 27361 & 0.557 &   9.7 & 0.9 &   $+$4.4 & 0.2 \\
16201 & 2014-07-21T22:13:45 & 56859.926 & 58390 & 0.894 &   7.0 & 0.4 &  $+$29.7 & 0.4 \\
16640 & 2014-07-24T11:21:26 & 56862.473 & 61679 & 0.789 &   6.6 & 0.4 &  $+$32.3 & 0.4 \\
16202 & 2014-08-19T15:30:01 & 56888.646 & 65128 & 0.935 &   4.3 & 0.3 &  $+$58.5 & 0.4 \\
17312 & 2014-08-22T06:21:18 & 56891.265 & 44895 & 0.939 &   3.8 & 0.3 &  $+$61.0 & 0.3 \\
16203 & 2014-09-02T12:47:11 & 56902.533 & 41423 & 0.946 &   4.5 & 0.4 &  $+$72.2 & 0.3 \\
17413 & 2014-09-08T15:21:28 & 56908.640 & 24650 & 0.931 &   3.9 & 0.5 &  $+$78.2 & 0.2 \\
17414 & 2014-09-13T12:24:59 & 56913.517 & 17317 & 0.942 &   3.2 & 0.5 &  $+$83.0 & 0.1 \\
16442 & 2014-10-25T13:38:44 & 56955.569 & 48350 & 0.943 &   2.8 & 0.3 & $+$125.3 & 0.3 \\
17545 & 2014-10-28T04:14:57 & 56958.177 & 34530 & 0.941 &   3.0 & 0.3 & $+$127.8 & 0.2 \\
 5906 & 2006-01-21T19:04:02 & 53756.794 & 12317 & 0.997 &   2.8 & 0.6 & $+$132.0 & 0.1 \\
17544 & 2014-11-01T16:52:08 & 56962.703 & 25642 & 0.942 &   2.8 & 0.4 & $+$132.3 & 0.2 \\
 7263 & 2006-01-22T16:51:51 & 53757.703 & 42528 & 0.997 &   3.2 & 0.3 & $+$133.1 & 0.3 \\
 7264 & 2006-01-30T15:06:27 & 53765.629 & 37593 & 1.000 &   3.2 & 0.3 & $+$141.0 & 0.2 \\
16443 & 2014-11-14T23:14:31 & 56975.968 & 34530 & 0.944 &   2.3 & 0.3 & $+$145.6 & 0.2 \\
17486 & 2014-12-04T13:39:50 & 56995.569 & 33541 & 0.941 &   2.3 & 0.3 & $+$165.2 & 0.2 \\
17555 & 2014-12-06T16:40:37 & 56997.695 & 55247 & 0.945 &   2.7 & 0.3 & $+$167.4 & 0.3 \\
17561 & 2014-12-20T17:22:40 & 57011.724 & 54567 & 0.945 &   2.5 & 0.2 & $+$181.5 & 0.3 \\
17562 & 2014-12-25T15:11:01 & 57016.633 & 42031 & 0.946 &   2.2 & 0.3 & $+$186.3 & 0.3 \\
16444 & 2014-12-27T22:58:58 & 57018.958 & 41440 & 0.942 &   2.7 & 0.3 & $+$188.6 & 0.3 \\
16448 & 2015-02-14T11:54:08 & 57067.496 & 34599 & 0.943 &   2.5 & 0.3 & $+$237.1 & 0.2 \\
17602 & 2015-02-19T13:57:46 & 57072.582 & 51705 & 0.944 &   2.4 & 0.2 & $+$242.3 & 0.3 \\
16447 & 2015-03-26T05:26:59 & 57107.227 & 26868 & 0.944 &   1.9 & 0.3 & $+$276.8 & 0.2 \\
16199 & 2015-03-27T20:27:05 & 57108.852 & 39461 & 0.943 &   2.2 & 0.3 & $+$278.5 & 0.2 \\
17640 & 2015-03-31T13:14:43 & 57112.552 & 26318 & 0.941 &   2.3 & 0.4 & $+$282.1 & 0.2 \\
17641 & 2015-04-04T19:45:40 & 57116.823 & 24638 & 0.939 &   2.2 & 0.4 & $+$286.4 & 0.2 \\
\hline
\end{tabular}
\end{table*}

\begin{figure}
\begin{center}
\includegraphics[width=\columnwidth]{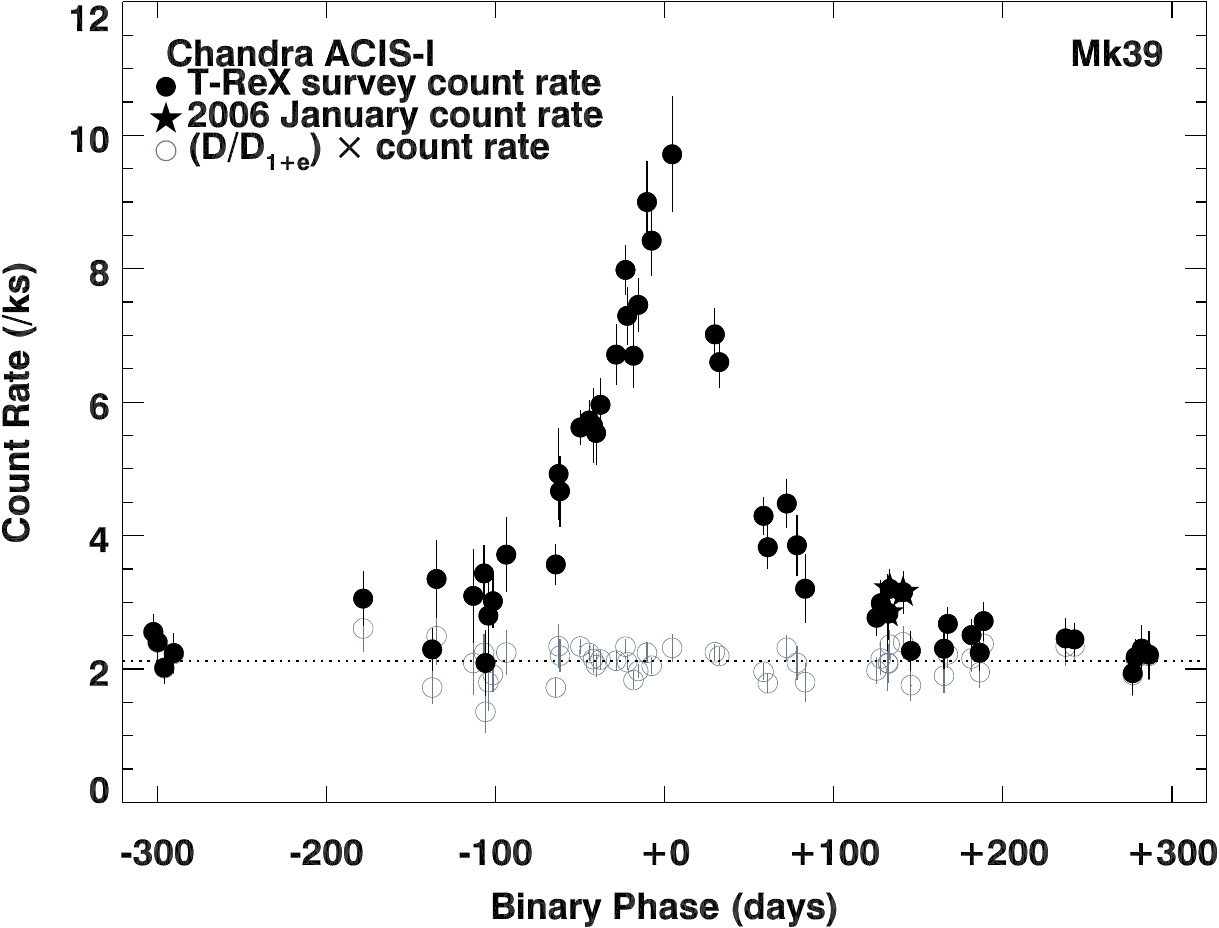}
\end{center}
\caption{The X-ray orbit of Mk~39 with $P = 641.1$~d and $T_0 = 56830.6$~MJD with observed T-ReX count rates in solid black and the same values adjusted for the implied inverse binary separation shown in open grey. \label{X-rayOrbit} }
\end{figure}

Figure~\ref{X-rayOrbit} shows the folded X-ray light curve resulting from the $1/D$ model of the observed T-ReX count rates logged in Table~\ref{ChandraLog}. Shown in grey are the
observed count rate values scaled by the inverse ratio of the implied binary separation to the maximum separation at apastron, $D_{1+e}$. In a successful model, this quantity would be
independent of phase. In this case, the observed dynamic range of about a factor of 5 is reproduced to within a few percent both before periastron and in the less well
sampled data after. Defects appear to be largely absent that could have been due to a variety of potential instrumental or physical causes such as uncertainties in the often substantial count-rate corrections shown in the table or the effects of absorption through stellar winds close to conjunctions.

For the purposes of predicting timing of future observations at or near periastron, in addition to the most accurately determined period of $P = 648.6\pm0.9$~d from the radial
velocity analysis,
it would be worth keeping in mind $P = 648.3\pm0.9$~d from the weighted mean of optical and X-ray values and $P = 647.6\pm1.4$~d from alignment of the putative photometric eclipse
discussed in section~\ref{properties}.



\bsp	
\label{lastpage}
\end{document}